\documentclass[superscriptaddress,showkeys,showpacs,preprintnumbers,twocolumn]{revtex4-1}

\usepackage{epsfig}
\usepackage{graphicx}
\usepackage{pslatex}
\usepackage{color}
\usepackage{amssymb}
\usepackage{xspace}

\def\kT{\ensuremath{k_{\rm T}}\xspace}
\def\mT{\ensuremath{m_{\rm T}}\xspace}

\begin{document}

\title{Freeze-out Dynamics via Charged Kaon Femtoscopy in $\sqrt{s_{NN}}$=200 GeV Central Au+Au Collisions}
%
\affiliation{AGH University of Science and Technology, Cracow, Poland}
\affiliation{Argonne National Laboratory, Argonne, Illinois 60439, USA}
\affiliation{University of Birmingham, Birmingham, United Kingdom}
\affiliation{Brookhaven National Laboratory, Upton, New York 11973, USA}
\affiliation{University of California, Berkeley, California 94720, USA}
\affiliation{University of California, Davis, California 95616, USA}
\affiliation{University of California, Los Angeles, California 90095, USA}
\affiliation{Universidade Estadual de Campinas, Sao Paulo, Brazil}
\affiliation{Central China Normal University (HZNU), Wuhan 430079, China}
\affiliation{University of Illinois at Chicago, Chicago, Illinois 60607, USA}
\affiliation{Cracow University of Technology, Cracow, Poland}
\affiliation{Creighton University, Omaha, Nebraska 68178, USA}
\affiliation{Czech Technical University in Prague, FNSPE, Prague, 115 19, Czech Republic}
\affiliation{Nuclear Physics Institute AS CR, 250 68 \v{R}e\v{z}/Prague, Czech Republic}
\affiliation{University of Frankfurt, Frankfurt, Germany}
\affiliation{Institute of Physics, Bhubaneswar 751005, India}
\affiliation{Indian Institute of Technology, Mumbai, India}
\affiliation{Indiana University, Bloomington, Indiana 47408, USA}
\affiliation{Alikhanov Institute for Theoretical and Experimental Physics, Moscow, Russia}
\affiliation{University of Jammu, Jammu 180001, India}
\affiliation{Joint Institute for Nuclear Research, Dubna, 141 980, Russia}
\affiliation{Kent State University, Kent, Ohio 44242, USA}
\affiliation{University of Kentucky, Lexington, Kentucky, 40506-0055, USA}
\affiliation{Institute of Modern Physics, Lanzhou, China}
\affiliation{Lawrence Berkeley National Laboratory, Berkeley, California 94720, USA}
\affiliation{Massachusetts Institute of Technology, Cambridge, MA 02139-4307, USA}
\affiliation{Max-Planck-Institut f\"ur Physik, Munich, Germany}
\affiliation{Michigan State University, East Lansing, Michigan 48824, USA}
\affiliation{Moscow Engineering Physics Institute, Moscow Russia}
\affiliation{National Institute of Science Education and Research, Bhubaneswar 751005, India}
\affiliation{Ohio State University, Columbus, Ohio 43210, USA}
\affiliation{Old Dominion University, Norfolk, VA, 23529, USA}
\affiliation{Institute of Nuclear Physics PAN, Cracow, Poland}
\affiliation{Panjab University, Chandigarh 160014, India}
\affiliation{Pennsylvania State University, University Park, Pennsylvania 16802, USA}
\affiliation{Institute of High Energy Physics, Protvino, Russia}
\affiliation{Purdue University, West Lafayette, Indiana 47907, USA}
\affiliation{Pusan National University, Pusan, Republic of Korea}
\affiliation{University of Rajasthan, Jaipur 302004, India}
\affiliation{Rice University, Houston, Texas 77251, USA}
\affiliation{Universidade de Sao Paulo, Sao Paulo, Brazil}
\affiliation{University of Science \& Technology of China, Hefei 230026, China}
\affiliation{Shandong University, Jinan, Shandong 250100, China}
\affiliation{Shanghai Institute of Applied Physics, Shanghai 201800, China}
\affiliation{SUBATECH, Nantes, France}
\affiliation{Temple University, Philadelphia, Pennsylvania, 19122, USA}
\affiliation{Texas A\&M University, College Station, Texas 77843, USA}
\affiliation{University of Texas, Austin, Texas 78712, USA}
\affiliation{University of Houston, Houston, TX, 77204, USA}
\affiliation{Tsinghua University, Beijing 100084, China}
\affiliation{United States Naval Academy, Annapolis, MD 21402, USA}
\affiliation{Valparaiso University, Valparaiso, Indiana 46383, USA}
\affiliation{Variable Energy Cyclotron Centre, Kolkata 700064, India}
\affiliation{Warsaw University of Technology, Warsaw, Poland}
\affiliation{University of Washington, Seattle, Washington 98195, USA}
\affiliation{Wayne State University, Detroit, Michigan 48201, USA}
\affiliation{Yale University, New Haven, Connecticut 06520, USA}
\affiliation{University of Zagreb, Zagreb, HR-10002, Croatia}

\author{L.~Adamczyk}\affiliation{AGH University of Science and Technology, Cracow, Poland}
\author{J.~K.~Adkins}\affiliation{University of Kentucky, Lexington, Kentucky, 40506-0055, USA}
\author{G.~Agakishiev}\affiliation{Joint Institute for Nuclear Research, Dubna, 141 980, Russia}
\author{M.~M.~Aggarwal}\affiliation{Panjab University, Chandigarh 160014, India}
\author{Z.~Ahammed}\affiliation{Variable Energy Cyclotron Centre, Kolkata 700064, India}
\author{I.~Alekseev}\affiliation{Alikhanov Institute for Theoretical and Experimental Physics, Moscow, Russia}
\author{J.~Alford}\affiliation{Kent State University, Kent, Ohio 44242, USA}
\author{C.~D.~Anson}\affiliation{Ohio State University, Columbus, Ohio 43210, USA}
\author{A.~Aparin}\affiliation{Joint Institute for Nuclear Research, Dubna, 141 980, Russia}
\author{D.~Arkhipkin}\affiliation{Brookhaven National Laboratory, Upton, New York 11973, USA}
% ...

\collaboration{STAR Collaboration}\noaffiliation

\date{\today}

\begin{abstract}
We present measurements of three-dimensional correlation functions of like-sign low transverse momentum kaon pairs from $\sqrt{s_{NN}}$=200~GeV Au+Au collisions.
A Cartesian surface-spherical harmonic decomposition technique was used to extract the kaon source function. 
The latter was found to have a three-dimensional Gaussian shape and can be adequately reproduced by Therminator event generator simulations with resonance contributions taken into account.
Compared to the pion one, the kaon source function is generally narrower and does not have the long tail along the pair transverse momentum direction.
The kaon Gaussian radii display a monotonic decrease with increasing transverse mass \mT over the interval of 0.55$\le$\mT{}$\le$1.15 GeV/$c^{2}$. 
While the kaon radii are adequately described by the \mT{}-scaling in the outward and sideward directions, in the longitudinal direction the  lowest \mT  value exceeds the expectations from a pure hydrodynamical model prediction.


\end{abstract}

\maketitle

\section{Introduction}

Analysis of the data collected at the Relativistic Heavy Ion Collider (RHIC) has resulted in the discovery of strongly interacting, almost perfect fluid created in high energy nucleus-nucleus collisions~\cite{brahmsw, phobosw, starw, phenixw}. 
Lattice calculations predict that the transition between normal nuclear matter and this new phase is a smooth crossover~\cite{aok06}. This is consistent with the absence of long source lifetimes which would indicate a first-order phase transition \cite{lis05}.
Moreover, analysis of three-dimensional (3D) two-pion correlation functions, exploiting the novel technique of Cartesian surface-spherical harmonic decomposition of Danielewicz and Pratt \cite{dan05,dan06}, revealed significant non-Gaussian features in the pion source function \cite{chu08}. Furthermore, the extraction of the shape of the pion source function in conjunction with model comparisons has permitted the decoupling of the spatio-temporal observable into its spatial and temporal aspects, and the latter into source lifetime and emission duration. However, an interpretation of pion correlations in terms of pure hydrodynamic evolution is complicated by the significant contributions of resonance decays. A purer probe of the fireball decay could be obtained
with kaons which suffer less contribution from long lifetime resonances and have a smaller
rescattering cross-section than pions. 
The lower yields, however, make it difficult to carry out a detailed 3D source shape analysis of kaons.
A 1D kaon source image measurement was recently
reported by the PHENIX Collaboration \cite{aki09}. This measurement, 
however, corresponds to a fairly broad range of the pair transverse momentum $2\kT$  
which makes the interpretation more ambiguous. In particular, 
information about the transverse expansion of the system, contained in the \kT
dependence of the emission radii, is lost. The 1D nature of the
measurement has also less constraining power on model predictions 
than would be available from a 3D measurement.

This paper presents a full 3D analysis of the correlation function of midrapidity, low transverse momentum like--sign kaon pairs. 
The technique used in this paper is similar to that employed in the first 3D extraction of the pion source function \cite{chu08}.
It involves the decomposition of the 3D kaon correlation function into a basis of Cartesian
surface-spherical harmonics to yield coefficients, also called moments, of the decomposition which are then fitted with a trial functional form for the 3D source function. The latter is then compared to models to infer the dynamics behind the fireball expansion.

\section{Experiment and Datasets}

The presented data from Au$+$Au collisions at $\sqrt {s_{NN}}= 200$ GeV were 
taken by the STAR Collaboration during the year-2004 and 2007 runs. A total of 4.6 million 0-20\% central events were used from year 2004, and 16 million 0-20\% central events from year 2007.
We also analyzed 6.6 million 0--30\% central events from the 
year 2004 run to compare to the previously published PHENIX 
kaon measurements \cite{aki09}.
Charged tracks are detected in the STAR Time Projection Chamber (TPC)
\cite{stardet}, surrounded by a solenoidal magnet providing a nearly uniform magnetic field of 0.5 T along
the beam direction. The TPC is used both for the tracking  of charged particles
at midrapidity  and particle identification by means of ionization energy loss. The $z$ position of the event vertex is constrained to be $|z|<30$~cm.

\section{Source Shape Analysis}

\subsection{Correlation moments}

The 3D correlation function $C(\mathbf{q}) = N_{\rm {same}}(\mathbf{q})/N_{\rm {mixed}}
(\mathbf{q})$
is constructed as the ratio of the 3D relative momentum distribution, 
$N_{\rm {same}}(\mathbf{q})$, for $K^+K^+$ and $K^-K^-$ pairs in the same event
to that from mixed events, $N_{\rm {mixed}}(\mathbf{q})$. Here,
$\mathbf{q}=(\mathbf{p_1}-\mathbf{p_2})/2$, where $\mathbf{p_1}$ 
and $\mathbf{p_2}$ are the momentum 3-vectors of the particles in the 
pair center-of-mass system (PCMS). 
The non-commutativity of the Lorentz transformations along non--collinear directions demands that the Lorentz transformation from the laboratory frame to the PCMS is made by first transforming to the pair longitudinally co-moving system (LCMS) along the beam direction and then to the PCMS along the pair transverse momentum.
$C(\mathbf{q})$ is flat and normalized to unity over $60<|\mathbf q|<100$~MeV/$c$.

To obtain the moments, the 3D correlation function
$C(\mathbf{q})$, is expanded 
in a 
Cartesian harmonic basis~\cite{dan05,dan06}
\begin{equation}
C(\mathbf{q})-1 \equiv R(\mathbf{q}) =\sum_{l, \alpha_1 \ldots
\alpha_l}R^l_{\alpha_1
 \ldots \alpha_l}(q) \,A^l_{\alpha_1 \ldots \alpha_l} (\Omega_\mathbf{q})\ ,
\label{eqn1}
\end{equation}
where $l=0,1,2,\ldots$, $\alpha_i=x, y \mbox{ or } z$, and
$A^l_{\alpha_1 \ldots \alpha_l}(\Omega_\mathbf{q})$
are Cartesian harmonic basis elements ($\Omega_\mathbf{q}$ is the solid 
angle in $\mathbf{q}$ space). $R^l_{\alpha_1 \ldots \alpha_l}(q)$, where $q$ is the modulus of $\mathbf q$, are Cartesian correlation moments,
\begin{equation}
 R^l_{\alpha_1 \ldots \alpha_l}(q) = \frac{(2l+1)!!}{l!}
 \int \frac{d \Omega_\mathbf{q}}{4\pi} A^l_{\alpha_1 \ldots \alpha_l} 
 (\Omega_\mathbf{q}) \, R(\mathbf{q}).
 \label{eqn2}
\end{equation}
The coordinate axes $x$-$y$-$z$ form a right-handed \textit{out-side-long} Cartesian
coordinate system. They are oriented so that the $z$-axis is parallel to the beam
direction and $x$ points in the direction of the pair total transverse momentum.

Correlation moments can be calculated from the 
measured 3D correlation function using Eq.~(\ref{eqn2}). 
Even moments with $l>4$ were found to be consistent with zero 
within statistical uncertainty. As expected from symmetry considerations,
the same was also found for odd moments.  
Therefore in this analysis, the sum in Eq.~(\ref{eqn1}) is truncated at $l=4$ and expressed 
in terms of independent moments only. 
Up to order 4, there are 6 independent moments: $R^0$, $R^2_{xx}$, $R^2_{yy}$, $R^4_{xxxx}$, $R^4_{yyyy}$ and $R^4_{xxyy}$.
Dependent moments are obtained from independent ones~\cite{dan05,dan06}.

These independent moments were extracted as a function of $q$, by fitting the truncated series to the measured 3D correlation function with the moments as free
parameters of the fit. 
The statistical errors on the moments reflect the statistical error on
the 3D correlation function. In order to estimate the effect of systematic errors, the 3D correlation function and associated moments were obtained under varying conditions including nominal vs. reverse magnetic field, year 2004 vs. year 2007 data, positively vs. negatively charged kaon pairs and varying
kaon sample purities. Although the variations did not introduce any observable systematic deviation in the correlation moments, they have some effect on the parameters of the 3D Gaussian  fit of Eq.~(\ref{ellip_eqn}).

\begin{figure}[!t]
 
 \vskip -1.0cm
 \includegraphics[width=\linewidth]{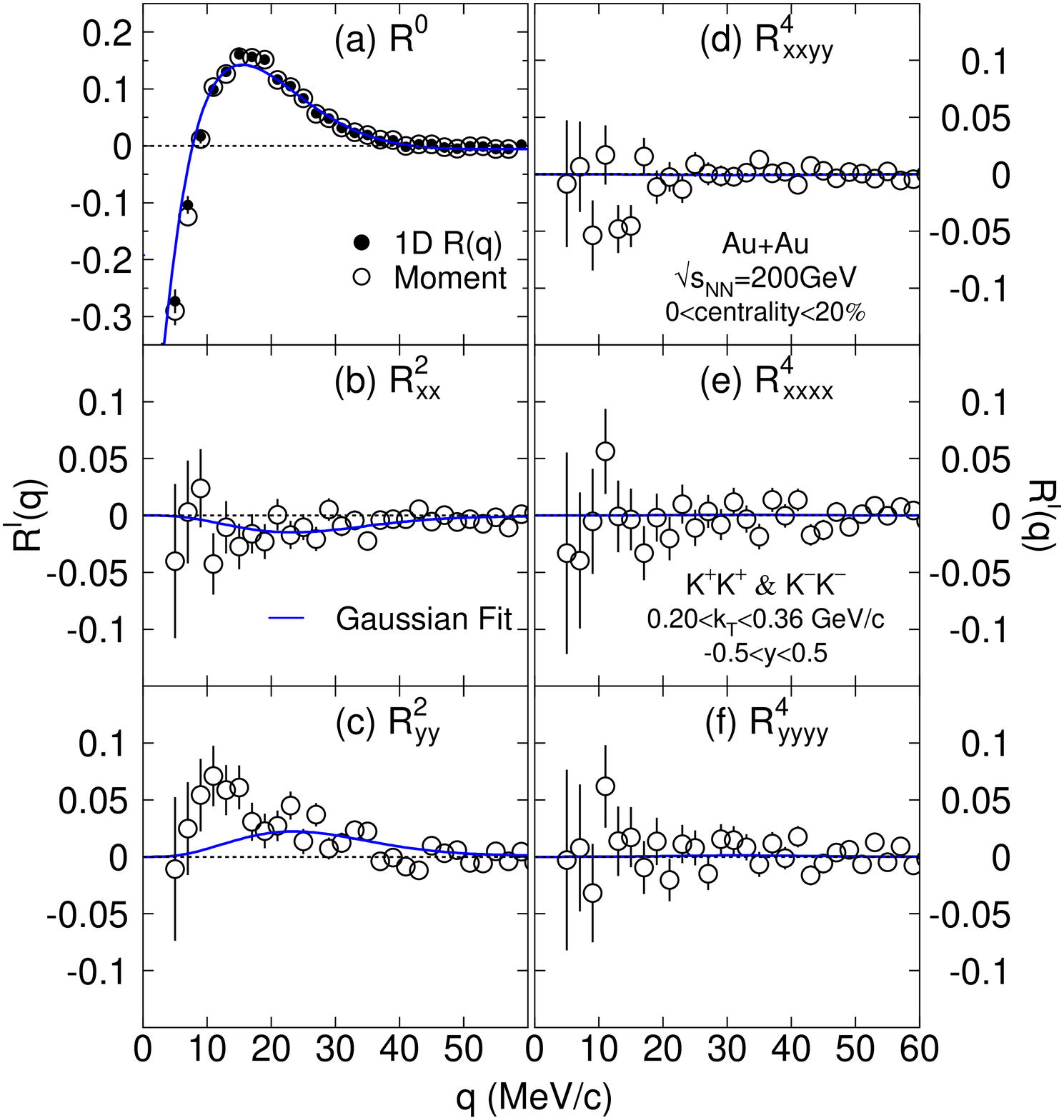}  
\vskip -1.0cm
   \caption{\label{kk_ggg_mom} {\it (Color online)} Independent correlation moments $R^l(q)$ for orders $l=0, 2, 4$ for midrapidity, low transverse momentum kaon pairs from the 20\% most central Au+Au collisions at $\sqrt{s_{NN}}$=200~GeV. Panel (a) also shows a comparison between $R^0(q)$ and $R(q)$. The error bars are statistical. The solid curves represent results of the Gaussian fit.}
\end{figure}

Figure~\ref{kk_ggg_mom} shows the independent correlation moments $R^l_{\alpha_1 \ldots
\alpha_l}$ up to order $l$=4 (open circles) for midrapidity ($|y|${}$<$0.5), low \kT
(0.2$<$\kT{}$<$0.36~GeV/$c$) kaon pairs produced in the 20\% most central 
Au+Au collisions at $\sqrt{s_{NN}}$=200~GeV; \kT is half the transverse momentum of the pair.
In panel (a), $R^0(q)$ is shown along with
the 1D correlation function $R(q) = C(q)-1$ (solid circles); both
represent angle-averaged correlation functions, but $R^0(q)$ is obtained from
the 3D correlation function via Eq.~(\ref{eqn2}) while $R(q)$ is
evaluated directly from the 1D correlation function. The data
points have been corrected for the effect of track momentum resolution. 
The agreement between $R^0(q)$ and $R(q)$  attests to the reliability of the
moment extraction technique.
Figures~\ref{kk_ggg_mom}(b)--(f) show that while second moments are already relatively small compared to their errors, fourth moments are insignificant without any visible trend. This further justifies truncating Eq.~(\ref{eqn1}) at $l=4$. 

\subsection{The 3D source function}

The probability of emitting a pair of particles with a pair separation vector $\mathbf{r}$ in the PCMS is given by the 3D source function $S(\mathbf{r})$. It is related to
the 3D correlation function $C(\mathbf{q})$ via a convolution integral ~\cite{lis05,Lednicky:2005tb} as
\begin{equation}
  C(\mathbf{q})-1 \equiv R(\mathbf{q}) = \int \left( |\phi(\mathbf{q},\mathbf{r})|^{2} -1 \right) S(\mathbf{r}) d\mathbf{r},
  \label{3dkpeqn}
\end{equation}
where the relative wave function $\phi(\mathbf{q},\mathbf{r})$ serves as a six-dimensional kernel, which in our case incorporates Coulomb interactions and Bose-Einstein symmetrization only~\cite{dan06}. Strong final state interactions are assumed to be negligible owing to the small $s$-wave scattering length ($\sim$0.1 fm) of two identical kaons~\cite{Abelev:2006gu}. Hence, no correction to the measured correlation function for Coulomb and other final-state interaction effects is required.
Analogously to Eq.~(\ref{eqn1}), the source function can be expanded in Cartesian harmonics basis elements as 
$S(\mathbf{r}) =\sum_{l, \alpha_1 \ldots
\alpha_l}S^l_{\alpha_1 \ldots \alpha_l}(r) A^l_{\alpha_1 \ldots \alpha_l} (\Omega_\mathbf{r})$.
Equation~(\ref{3dkpeqn}) can then be rewritten in terms of the independent moments~\cite{dan05,dan06}.

The 3D source function can be extracted by directly fitting 
the 3D correlation function with a trial functional form for 
$S(\mathbf r)$.  Because the 3D correlation function has been
decomposed into its
independent moments, this corresponds to a simultaneous fit of the 
six independent moments with the trial functional form.  
A four-parameter fit to the independent moments with a 3D Gaussian trial function,
\begin{eqnarray}
  S^G(r_x,r_y,r_z) = \frac{\lambda}{ {(2\sqrt\pi)}^3 R_x R_y R_z}
  \exp[-(\frac{r_x^2}{4 R_x^2} + \frac{r_y^2}{4 R_y^2} + \frac{r_z^2}{4 R_z^2})], \nonumber \\   
  \label{ellip_eqn}
\end{eqnarray}
yields a $\chi^2/\mathit{ndf}=1.7$. 
The correlation strength parameter $\lambda$ represents the integral short-distance contribution to the source function~\cite{Lednicky:1979ig}.
Figure~\ref{kk_ggg_mom} shows the fit as solid curves, making it evident that the quality of the fit is predominantly driven by the relatively small errors of $R^{0}(q)$.
The values of the Gaussian radii and the amplitude ($R_x, R_y, R_z$, $\lambda$) are listed in Table~\ref{tab:fits}.

\begin{figure}[!t]
 
 \vskip -1.0cm
 \includegraphics[width=\linewidth]{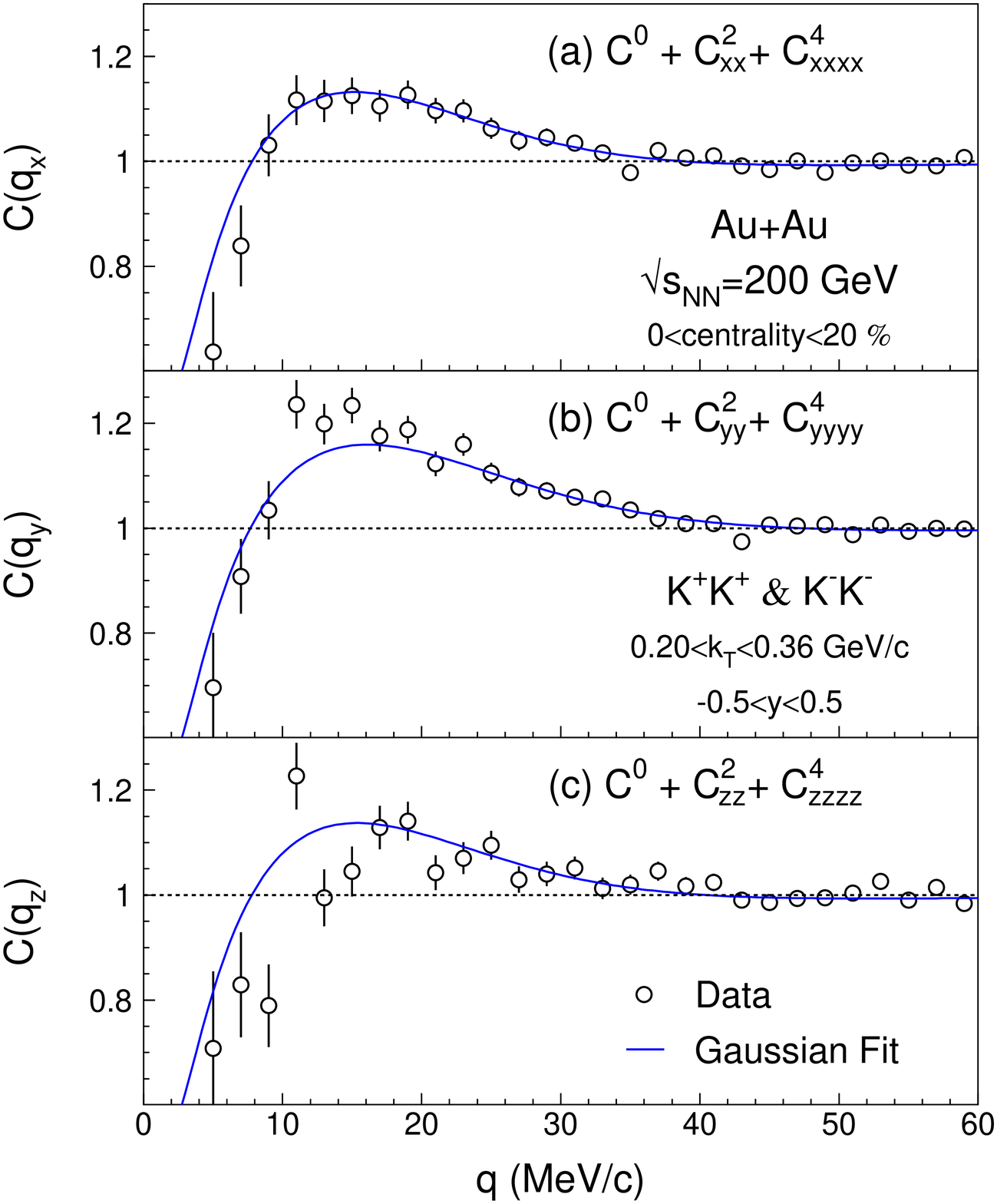} 
 \vskip -1.0cm
 \caption{\label{kk_ggg_cor} {\it (Color online)}
    {Kaon correlation function profiles (circles) for midrapidity, low transverse momentum kaon pairs from the 20\% most central Au+Au collisions at $\sqrt{s_{NN}}$=200~GeV (a) $C(q_x) \equiv
C(q_x,0,0)$, (b) $C(q_y) \equiv C(0,q_y,0)$ and (c) $C(q_z) \equiv C(0,0,q_z)$
in the $x$, $y$ and $z$ directions. The curves denote the Gaussian fit profiles.
}
  }
\end{figure}

Figure~\ref{kk_ggg_cor}(a)--(c) illustrate the kaon correlation function
profiles (circles) in the $x$, $y$ and $z$ directions ($C(q_x) \equiv C(q_x,0,0)$,
$C(q_y) \equiv C(0,q_y,0)$ and $C(q_z) \equiv C(0,0,q_z)$), respectively, obtained by summation of the relevant correlation terms 
$C^l_{\alpha_1 \ldots \alpha_l}(\mathbf{q})= \delta_{l,0} + R^l_{\alpha_1 \ldots \alpha_l}(q) A^l_{\alpha_1 \ldots \alpha_l} (\Omega_\mathbf{q})$ up to order $l$=4.
The peak at $q${}$\approx$20~MeV/$c$ is coming from an expected interplay of Coulomb repulsion at $q${}$\rightarrow$0 and Bose-Einstein enhancement.
The correlation profiles from the
data are well represented by the corresponding correlation profiles from the
Gaussian fit (line). Hence, the trial Gaussian shape for the kaon source
function seems to capture the essential components of the actual source
function.  

  \begin{figure}[!t]
\vskip -1.cm
\includegraphics[width=\linewidth]{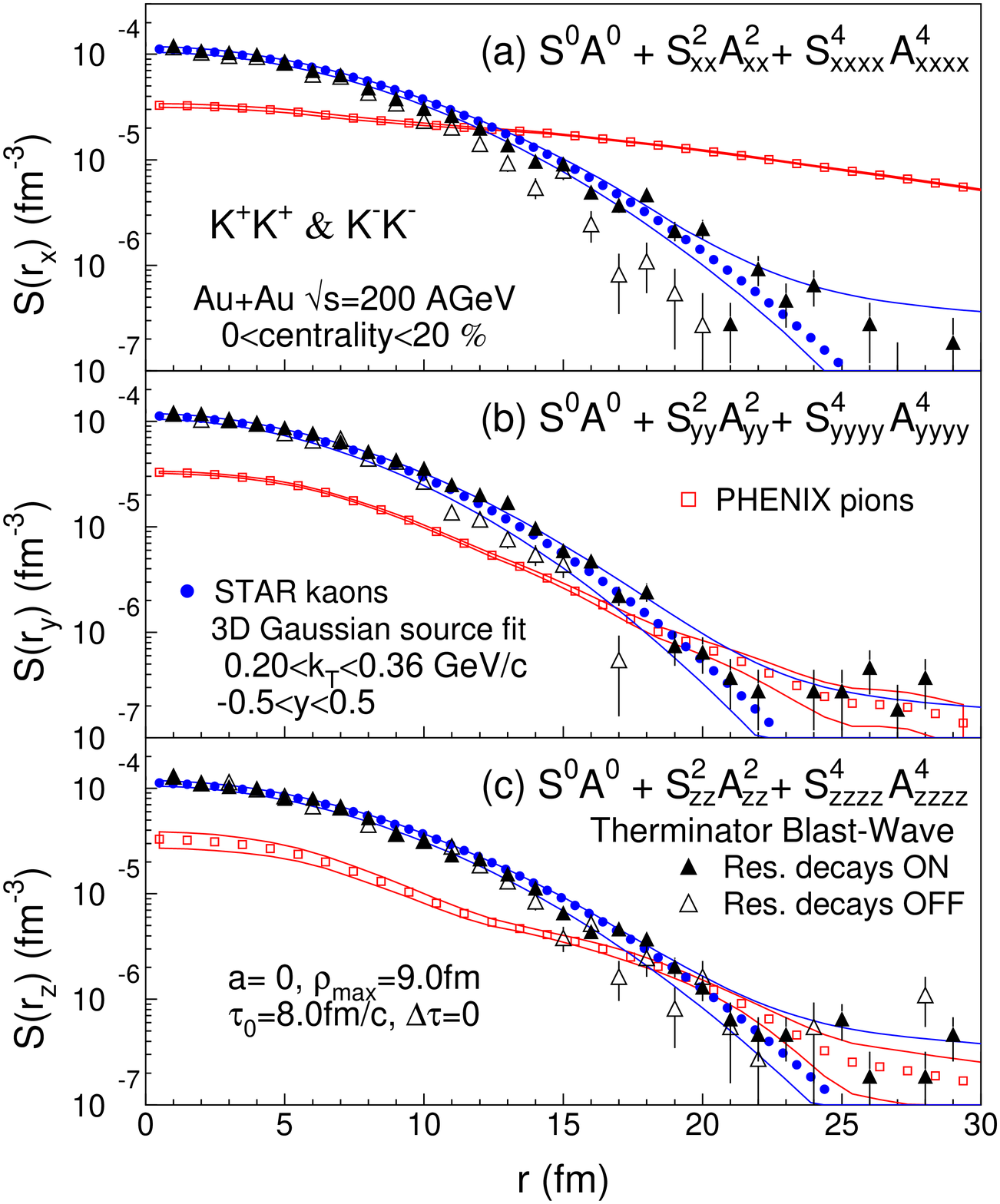}  
\vskip -1.cm
  \caption{\label{kk_ggg_src} {\it (Color online)}
    {Kaon source function profiles extracted from the data 
    (solid circles with error band) and 3D pion source function (squares) from PHENIX
\cite{chu08} together with Therminator model calculation for kaons with indicated parameter values
(triangles)}.
  }
\end{figure}

Figure~\ref{kk_ggg_src}(a)--(c) depict the extracted source function profiles in
the $x$, $y$ and $z$ directions ($S(r_x) \equiv S(r_x,0,0)$, $S(r_y) \equiv
S(0,r_y,0)$ and $S(r_z) \equiv S(0,0,r_z)$) obtained via the 3D
Gaussian fit (dots) to the correlation moments. The two solid curves around 
the Gaussian source function profiles 
represent the error band arising from the statistical and systematic 
errors on the 3D Gaussian fit parameters, as well as the 
uncertainty from the source shape assumption estimated using a 
double-Gaussian fit. Note that the latter  becomes important for large $r$ values only.

\subsection{Expansion dynamics and model comparison}

The source function profile $S(r_y)$ in the \textit{side} direction reflects the
mean transverse geometric size of the emission source, while the source lifetime
determines the extent of the source function profile $S(r_z)$ in the
\textit{long} direction.  
Being in the direction of the total pair transverse momentum (hence the
direction of Lorentz boost from the LCMS to PCMS frame), the source function
profile in the \textit{out} direction $S(r_x)$ is characterized by the
kinematic Lorentz boost, mean transverse geometric size as
well as source lifetime and particle emission duration. To disentangle
these various contributions, the Monte Carlo event generator Therminator
\cite{kis05} is used
to simulate the source breakup and emission dynamics.

The basic ingredients of the Therminator model employed in the analysis are (1) Bjorken assumption of longitudinal boost invariance; (2) blast-wave (BW) expansion in the transverse
direction with transverse velocity profile semilinear in transverse radius
$\rho$~\cite{kis07}, $v_r(\rho)=(\rho/\rho_{\rm max})/(\rho/\rho_{\rm max}+
v_t)$, where $v_t$=0.445 is obtained from BW fits to particle
spectra~\cite{abe09}; (3) after a proper lifetime $\tau$, a thermal emission of
particles takes place from the source elements distributed in a cylinder of infinite
longitudinal size and finite transverse dimension $\rho_{\rm max}$.
At the point of source breakup, all particle emission is
 collectively viewed as happening from a freeze-out
 hypersurface defined in the $\rho$-$\tau$~plane as $\tau = \tau_0 + a \rho$. Hence,
particles which are emitted from a generic source
 element with coordinates ($z$,$\rho$) will have emission time $t$ in the
laboratory frame given by $t^2 = (\tau_0 + a \rho)^2 + z^2$.

Note that the BW mode of fireball expansion means that $a=0$ \cite{kis06} making $\tau$ independent of $\rho$. Each source element is thus defined by only one value of the  proper breakup time $\tau=\tau_{0}$ and all particle emission from this  source element happens instantaneously in the rest frame of the  source element and the proper emission duration $\Delta\tau$ is set to 0. Later, we also discuss another choice for parameter $a$ which was used to describe the pion data \cite{chu08}.
 
Using a set of thermodynamic parameters previously tuned to fit charged pion and
kaon spectra~\cite{kis06}, midrapidity kaon pairs at low \kT were obtained from
Therminator with all known resonance decay processes on and off. They were then
boosted to the PCMS to obtain source function profiles for comparison with
corresponding profiles from the data.

Figures~\ref{kk_ggg_src}(a)--(c) indicate that the 3D source
function generated by the
Therminator model in the BW mode (solid triangles) with $\tau_0$=8.0$\pm$0.5~fm/$c$,
$\rho_{\rm max}$=9.0$\pm$0.5~fm and other previously tuned
parameters~\cite{abe09,kis06}, reproduces the experimentally extracted source
function profiles $S(r_x)$, $S(r_y)$ and $S(r_z)$. The calculations also show
that the source function excluding the contribution of resonances (open
triangles) is narrower than the experimentally observed Gaussian.
However, they do not allow us to draw a firm conclusion concerning the value of parameter $a$.  
Besides the Therminator default $a$=$0$, we tested the value $a$=$-0.5$, the same as used in Ref.~\cite{chu08} to describe the pion data. Our simulations with $a$=$-0.5$ and the other parameters fixed, underestimate  the source function $S(r_z)$ already for $r${}$>$5~fm but do not show any change in $S(r_x)$ and $S(r_y)$.  
Substantial improvement can be achieved if we allow at the same time $\tau_{0}$ to increase 
from 8 to $\sim$10.5~fm/$c$. The latter value is, however, considerably bigger than $\tau_{0}$=$8.5$ fm/c reported in \cite{chu08} for the pions.  Given these uncertainties, the scenario when kaon freeze-out occurs in the source element rest frame from a hypersurface devoid of any space-time correlation ($a$=$0$) is only marginally favored over the one where the emission occurs from the outer surface of the fireball inwards ($a${}$<$0).

Although most of the extracted parameters of the expanding fireball 
are consistent with those obtained from two-pion interferometry \cite{chu08}, the 3D source function shapes for kaons and  pions are very different. This is illustrated in Figs~\ref{kk_ggg_src} (a)--(c) which compares the correlation profiles for midrapidity
kaons (circles) with those for midrapidity pions (squares) reported by the
PHENIX Collaboration  \cite{chu08} for the same event centrality and transverse
momentum selection. The kaon source function profiles are
generally narrower in width  than those for pions.  Moreover, in contrast to the case for pions, a long tail is not observed in the kaon $S(r_x)$ (i.e. along the pair's total transverse momentum). Compared to the pion case  where a prominent cloud of resonance decay pions determines the source-function tail profiles in \textit{out} and \textit{long} directions \cite{chu08}, the narrower shape observed for the kaons indicates a much smaller role of long-lived resonance decays and/or of the exponential emission duration width $\Delta \tau$ on kaon emission.

\begin{figure}[!t]
  \vskip -1.cm
   \includegraphics[width=1.\linewidth, height=1.3\linewidth]{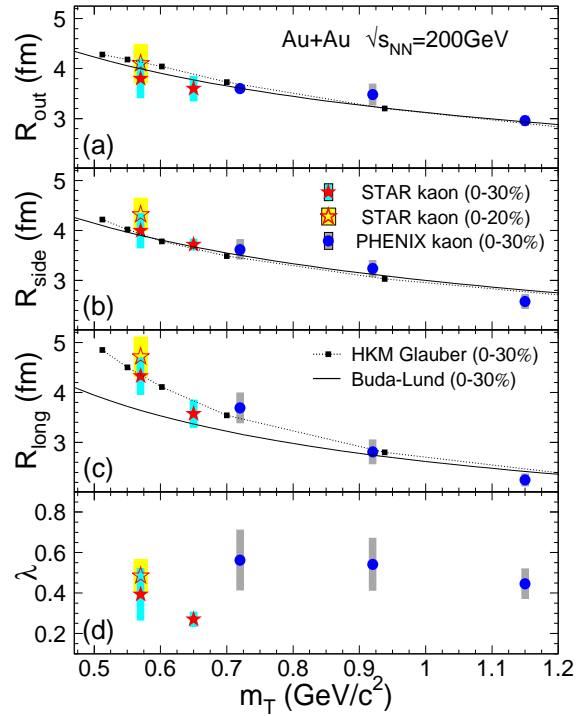}  
  \vskip -.8cm
  \caption{\label{gaus_radii_mt}  {\it (Color online)} Transverse mass dependence of Gaussian radii (a) $R_{out}$, (b) $R_{side}$ and (c) $R_{long}$   for midrapidity kaon pairs from the 30\%  most central  Au+Au collisions at $\sqrt{s_{NN}}$=200~GeV.  STAR data are shown as solid stars; PHENIX data \cite{aki09} as solid circles (error bars include both statistical and systematic uncertainties). Hydro-kinetic model \cite{kar10} with initial Glauber condition and Buda-Lund model \cite{Csanad:2008gt} calculations are shown by solid squares and solid curves, respectively. The dotted line between the solid squares is to guide the eye. For comparison purposes, we also plot the result from the 20\% most central Au+Au collisions as open stars. Panel (d) shows corresponding experimental values of the Gaussian fit parameter $\lambda$. }
  
\end{figure}

\section{\kT--{}Dependence}

Further insight into expanding fireball dynamics can be obtained by studying the \kT{}--dependence of the kaon Gaussian radii in LCMS. To achieve this goal, in addition to the lowest momentum bin (0.2$<$\kT{}$<$0.36 GeV/$c$), we have also analyzed  the kaon pairs with  0.36$<$\kT{}$<$0.48~GeV/$c$. The analysis was carried out for the 30\% most central Au+Au collisions at $\sqrt{s_{NN}}$=200~GeV. This wider centrality cut enabled us to compare our results to the PHENIX kaon data points obtained at higher \kT but at the same centrality \cite{aki09}.  
A 4--parameter fit to the two sets of independent moments with a Gaussian function Eq.~(\ref{ellip_eqn}) yields a $\chi^2/\mathit{ndf}$ of 1.1 and 1.3 respectively. The three Gaussian radii and the amplitude obtained from this fit are listed in Table~\ref{tab:fits}.
Note that the overall normalization of $S^G(r_x,r_y,r_z)$ may also be affected by systematic factors not included in this fit. While the value of $\lambda$ for the 0--20\% centrality data is only marginally smaller than that of Ref.~\cite{aki09}, the analysis of the 30\% most central collisions restricted to year 2004 data uses looser purity cuts, thus yielding substantially smaller $\lambda$. Additional dilution of the correlation strength  is expected from the $\phi \rightarrow K^+ K^-$ decays, which is, however, limited by the $\phi$ decay length of $\sim$11 fm in PCMS~\cite{Lednicky:1992me}. Calculations based on the core--halo model \cite{Vance:1998wd} employing the STAR $\phi/K^-$ ratio \cite{Adams:2004ux} yields a maximum 15--20\% decrease in $\lambda$ at low transverse momenta. Neither of those two effects  has a significant impact on the values of the extracted Gaussian radii.
 
Figure~\ref{gaus_radii_mt} shows the dependence of the Gaussian radii in LCMS ($R_{out}$=$R_{x}/\gamma$, ~$R_{side}$=$R_{y}$ and $R_{long}$=$R_{z}$; $\gamma$ is the kinematic Lorentz boost in the outward direction from the LCMS  to the PCMS frame) as a function of transverse mass $\mT=(m^{2}+\kT^{2})^{1/2}$ obtained from the fits to the 3D correlation functions
from STAR data (stars). The error bars on the STAR data are dominated by systematic uncertainties from particle identification and momentum resolution. The Gaussian radii for PHENIX kaon data \cite{aki09} (solid circles) are also shown, with the error bars representing statistical and systematic uncertainties combined. The model calculations from the Buda-Lund model~\cite{Csanad:2008gt} and from the hydrokinetic model (HKM)~\cite{kar10} are shown as solid curves and solid squares, respectively.
While the HKM provides a full microscopic transport simulation of hydrodynamic expansion of the system followed by dynamic decoupling, the Buda-Lund model is a pure analytical solution of the perfect fluid hydrodynamics. The latter describes the Gaussian radii of charged pions from Au+Au collisions~\cite{Adler:2004rq} at the same energy and centrality as our kaon data over the whole 0.30$\le$\mT{}$\le$1.15~GeV/$c^{2}$ interval~\cite{Csanad:2008gt}. Because the exact \mT{}--scaling is an inherent feature of perfect fluid hydrodynamics, the Buda-Lund model predicts that the kaon and pion radii fall on the same curve.

\begin{table}
\caption{\label{tab:fits} Parameters obtained from the 3D Gaussian source function fits for the different datasets. The first errors are statistical, the second errors are systematic.}
\begin{tabular*}{\columnwidth}{l @{\extracolsep{\fill}} c @{\extracolsep{\fill}}c @{\extracolsep{\fill}} c}
\hline\hline
Year & 2004+2007   & \multicolumn{2}{c}{2004} \\
Centrality  & 0\%--20\% & \multicolumn{2}{c}{0\%--30\%} \\ \cline{3-4}
\kT	[GeV/$c$]   & 0.2--0.36 & 0.2--0.36 & 0.36--0.48 \\
\hline
$R_x$ [fm] & 4.8$\pm$0.1$\pm$0.2 & 4.3$\pm$0.1$\pm$0.4 & 4.5$\pm$0.2$\pm$0.3 \\
$R_y$ [fm] & 4.3$\pm$0.1$\pm$0.1 & 4.0$\pm$0.1$\pm$0.3 & 3.7$\pm$0.1$\pm$0.1 \\
$R_z$ [fm] & 4.7$\pm$0.1$\pm$0.2 & 4.3$\pm$0.2$\pm$0.4 & 3.6$\pm$0.2$\pm$0.3  \\
$\lambda$  & 0.49$\pm$0.02$\pm$0.05 & 0.39$\pm$0.01$\pm$0.09 & 0.27$\pm$0.01$\pm$0.04 \\
$\chi^2/\mathit{ndf}$& 497/289 & 316/283 & 367/283 \\
\hline\hline
\end{tabular*}
\end{table}

From Figure~\ref{gaus_radii_mt} it is seen that the Gaussian radii for the kaon source function display a monotonic decrease with increasing transverse mass \mT from the STAR data at low \mT to the PHENIX data at higher \mT, as do the model calculations of Buda-Lund and HKM. 
The Gaussian radii in the outward and sideward directions are adequately described by both models over the whole interval. However, there is a marked difference between the HKM and the Buda-Lund predictions in the longitudinal direction, with the deviation becoming prominent for \mT{}$<$0.7 GeV/$c^{2}$ where the new STAR data reside. 
Our measurement at 0.2$\le$\kT$\le$0.36 GeV/$c$ clearly 
favors the HKM model as more representative of the expansion dynamics of the fireball, despite the fact that the Buda-Lund model describes pion data in all three directions. 
Hence, exact \mT{}--scaling of the Gaussian radii in the longitudinal direction between kaons and pions observed at lower energies~\cite{Afanasiev:2002fv} is not supported by our measurements.

\section{Conclusions}

In summary the STAR Collaboration has extracted the 3D source function for
midrapidity, low transverse momentum kaon pairs from central Au+Au collisions
at $\sqrt{s_{NN}}$=200~GeV via the method of Cartesian surface-spherical harmonic
decomposition. The source function is essentially a 3D Gaussian 
in shape. Comparison with Therminator model calculations indicates that kaons
are emitted from a fireball whose transverse dimension and lifetime are
consistent with those extracted with two-pion interferometry. However, the 
3D source function shapes for kaons and  pions are very different.
The narrower shape observed for the kaons indicates a much smaller role of long-lived resonance 
decays and/or of the exponential emission duration width $\Delta \tau$ on kaon emission.
The Gaussian radii for the kaon source function display a monotonic decrease with increasing transverse mass \mT over the interval 0.55$\le$\mT{}$\le$1.15 GeV/$c^{2}$. In the outward and sideward directions, this decrease is adequately described by \mT{}--scaling. However, in the longitudinal direction, the scaling is broken. The results are in favor of the hydro-kinetic predictions~\cite{kar10} over pure hydrodynamical model calculations. 

\section*{Acknowledgements}

We thank the RHIC Operations Group and RCF at BNL, the NERSC Center at LBNL and the Open Science Grid consortium for providing resources and support. This work was supported in part by the Offices of NP and HEP within the U.S. DOE Office of Science, the U.S. NSF, the Sloan Foundation, CNRS/IN2P3; FAPESP CNPq of Brazil; Ministry of Education and Science of the Russian Federation; NNSFC, CAS, MoST, and MoE of China; GA and MSMT of the Czech Republic; FOM and NWO of the Netherlands; DAE, DST, and CSIR of India; Polish Ministry of Sci. and Higher Ed., National Research Foundation (NRF-2012004024); Ministry of Science, Education and Sports of the Republic of Croatia; and RosAtom of Russia.

\bibliographystyle{elsarticle-num}

\end{document}